\documentclass[aps,showpacs,twocolumn]{revtex4-1}
\usepackage[english]{babel}
\usepackage{graphicx}
\usepackage{color}

\begin{document}
\title{Properties of random sequential adsorption of generalized dimers}
\author{Micha\l{} Cie\'sla}
 \email{michal.ciesla@uj.edu.pl}
\affiliation{M. Smoluchowski Institute of Physics, Jagiellonian University, 30-059 Krak\'ow, Reymonta 4, Poland.}%
\date{\today}

\begin{abstract}
Saturated random packing of particles built of two identical, relatively shifted spheres in two and three dimensional flat and homogeneous space was studied numerically using random sequential adsorption algorithm. The shift between centers of spheres varied from $0.0$ to $8.0$ sphere diameters. Numerical simulations allowed determine random sequential adsorption kinetics, saturated random coverage ratio as well as available surface function and  density autocorrelation function.
\end{abstract}
\maketitle
\section{Introduction}
Adsorption at various interfaces underlies a number of extremely important processes of utilitarian significance for technology and science. For example, it is crucial for separation of mixtures, filtration and purification processes. Protein adsorption is involved in cell adhesion, inflammatory response, preventing artificial organ and biomaterial rejection, plaque formation and  fouling of contact lenses. It is also prerequisite for chromatography, filtration, biocatalysis, biosensing, immunological assays, etc. \cite{bib:Dabrowski2001}.
\par 
Since its introduction by Feder \cite{bib:Feder1980}, random sequential adsorption (RSA) became one of the major tools for modeling monolayers obtained during irreversible adsorption processes, at first for spherical adsorbates and, then, for more complex particles like squares, spherocylinders and needles \cite{bib:Talbot1989, bib:Vigil1989, bib:Tarjus1991, bib:Viot1992, bib:Ricci1992}. In recent studies RSA has also been successfully used for modeling complex particles using coarse-grain approximation \cite{bib:Rabe2011, bib:Finch2012, bib:Katira2012, bib:Adamczyk2012rev}. For example, a coarse-grain fibrinogen model successfully explains the density of adsorbed proteins for a wide range of experimental conditions \cite{bib:Adamczyk2010, bib:Adamczyk2011, bib:Ciesla2013fib}. 
\par
Random sequential adsorption of generalized dimers has been mentioned for the first time  in \cite{bib:Ciesla2013ring}, however, in the context of RSA kinetics only and just for the case of distance between disks centers lower than disk diameter. Aim of this study is to extend this analysis to determine all the main properties of obtained adsorption monolayers such as saturated random coverage ratio, density autocorrelations and available surface function (ASF), as well as to study adsorption of particles for which the distance between disks centers exceed disk diameter. Such particles can possibly model many biological molecules like flexible dimers or dumbbell-shape molecules \cite{bib:Ju2012,bib:Hu2013} and, what is even more important from the theoretical point of view, they have not been studied earlier.
\section{Model}
\begin{figure}[htb]
\centerline{%
\includegraphics[width=0.8\columnwidth]{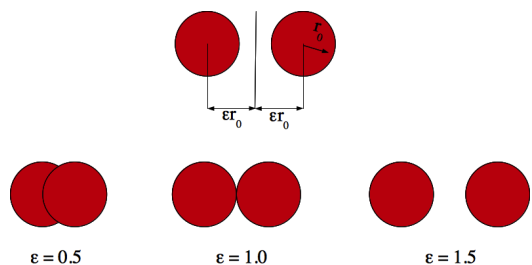}}
\caption{(Color online) Definition of anisotropy parameter $\epsilon$, and examples of three generalized dimers for $\epsilon=0.5$, $\epsilon=1.0$ and $\epsilon=1.5$.}
\label{fig:epsilon}
\end{figure}

A generalized dimer is build of two identical spheres of radius $r_0$ (see Fig.\ref{fig:epsilon}). Width-to-height ratio for such molecules can be expressed as $(1+\epsilon)$ where $\epsilon$ is the measure of spheres displacement. 
The case of $\epsilon=1$ has been extensively studied in \cite{bib:Ciesla2012dim,bib:Ciesla2013dim}. Dimers were placed on a square flat homogeneous collector surface of a side size $1000 \, r_0$ according to the RSA algorithm, which iteratively repeats the following steps:
\begin{description}
\item[-] A virtual generalized dimer of a fixed distance between spheres is created. Its position and orientation on a collector is chosen randomly according to the uniform probability distribution; however, centers of both the spheres are required to be on a collector.
\item[-] An overlapping test is performed for previously adsorbed nearest neighbors of the virtual molecule. The test checks if surface-to-surface distance between each of the spheres is greater than zero.
\item[-] If there is no overlap the virtual particle is irreversibly adsorbed and added to an existing covering layer. Its position does not change during further calculations.
\item[-] If there is an overlap the virtual dimer is removed and abandoned.
\end{description}
The number of RSA iterations $N$ is typically expressed in dimensionless time units:
\begin{equation}
\label{eq:dimlesstime}
t = N\frac{S_{\rm M}(\epsilon)}{S_{\rm C}},
\end{equation}
where
\begin{equation}
S_{\rm M}(\epsilon) = \left\{ 
  \begin{array}{cc} \left[ 2\pi - \left( \arccos \epsilon - \sqrt{1-\epsilon^2} \right) \right] r_0^2  & \mbox{if } \epsilon<1, \\
2 \pi r_0^2 & \mbox{ otherwise},
\end{array}
\right.
\end{equation}
is an area covered by a single dimer, and $S_{\rm C}=10^6 \, r_0^2$ is a collector surface area. Simulation is halted after the number of steps corresponding to $t = 10^5$. To improve statistics, $100$ independent simulations were performed for a single generalized dimer model. During the simulations, the coverage ratio $\theta$ has been measured:
\begin{equation}
\theta(t) = n(t) \frac{S_{\rm M}}{S_{\rm C}},
\end{equation}
where $n(t)$ is the number of adsorbed molecules after dimensionless time $t$.
\section{Results}
\begin{figure}[htb]
\vspace{1cm}
\centerline{%
\includegraphics[width=0.9\columnwidth]{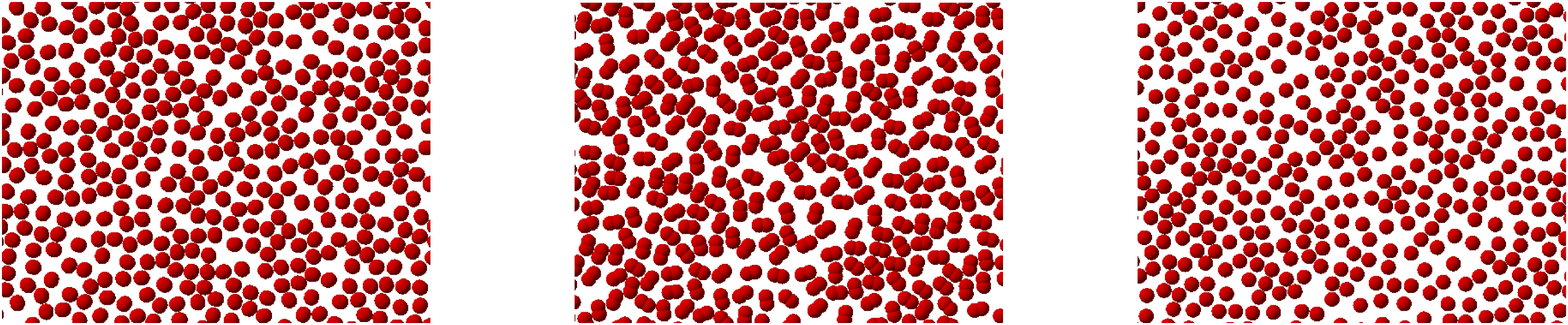}}
\caption{(Color online) Example coverages for three different values of $\epsilon$: $\epsilon=0.01$, $\epsilon=0.5$, and $\epsilon=1.4$ at the end of the simulation. The coverage ratio $\theta$ for presented snapshots are lower by about 0.5\% from saturated random coverage ratio $\theta_{\rm max}$}.
\label{fig:examples}
\end{figure}

Obtained example coverages are presented in Fig.\ref{fig:examples}.
The most important parameter which can be measured in numerical simulations is saturated random coverage ratio $\theta_{\rm max} \equiv \theta(t \to \infty )$, as it can be easily obtained experimentally and thus used for the comparison. However, simulations involve a finite number of RSA algorithm steps. Therefore, to determine $\theta_{\rm max}$, knowledge about RSA kinetics is essential.
\subsection{RSA kinetics}
As observed by Feder \cite{bib:Feder1980} and confirmed by later analitical studies \cite{bib:Pomeau1980, bib:Swendsen1981, bib:Privman1991}, the RSA kinetics for spheres adsorption on a flat collector is governed by the following power law:
\begin{equation}
\theta_{\rm max} - \theta(t) \sim t^{-1/d},
\label{fl}
\end{equation}
where $\theta(t)$ is a coverage ratio after time $t$ and $d$ is a collector dimension. The above relation, known as Feder's law, has been also proved numerically for one- to eight- dimensional collectors \cite{bib:Zhang2013} as well as for fractal collectors having $1<d<3$~\cite{bib:Ciesla2012b, bib:Ciesla2013b}. It is also valid for RSA of anisotropic particles on a flat collector \cite{bib:Ricci1992,bib:Adamczyk2010, bib:Ciesla2013tet,bib:Ciesla2013hex}, however, in this case, $d=3$, which could be explained by additional degree of freedom of an adsorbate particle \cite{bib:Viot1990, bib:Viot1992, bib:Hinrichsen1986, bib:Ciesla2013pol}.
\par
\begin{figure}[htb]
\vspace{1cm}
\centerline{%
\includegraphics[width=0.8\columnwidth]{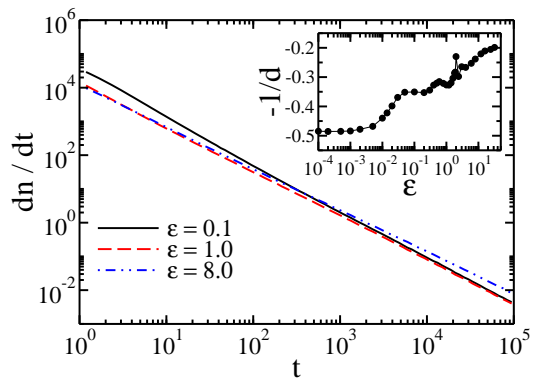}}
\caption{(Color online) The dependence of increments of the mean number of adsorbed particles on the dimensionless time for different anisotropy $\epsilon$. Inset shows the exponent in Eq.(\ref{fl}) determined using least square fit method.}
\label{fig:fl}
\end{figure}

In our case, the power law (\ref{fl}) is fulfilled for all studied particles (see Fig.\ref{fig:fl}).
For a very small anisotropy ($\epsilon<0.01$), the RSA kinetics is similar to the one observed for disks ($d \approx 2$). Then, transition to $d=3$ -- the characteristic value for anisotropic adsorbates -- occurs. Interestingly, when disks are separated ($\epsilon>1$), parameter $d$ increases up to $4.5$. Around $\epsilon =2$ maximum is observed; for such anisotropy, a disk from one generalized dimers can fit between disks belonging to other particle. On the one hand, it allows denser packing, but on the other hand it slows down the kinetics as it is hard to randomly place a particle in a such restricted area. 
\subsection{Saturated random coverage ratio}
\begin{figure}[htb]
\vspace{1cm}
\centerline{%
\includegraphics[width=0.8\columnwidth]{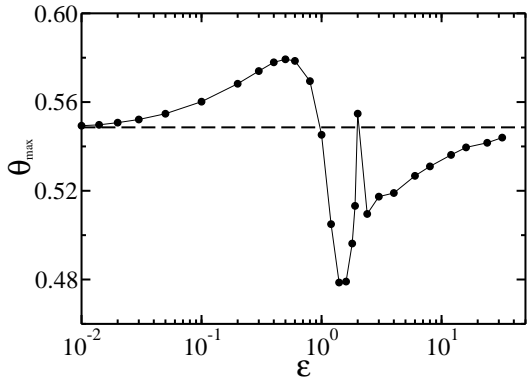}}
\caption{The dependence of saturated random coverage on dimer anisotropy. Dots correspond to measured data.  Statistical error of $\theta_{\rm max}$ is approximately $0.0001$ and is smaller than the  dot size. Lines have been drawn to guide the eye. Dashed line corresponds to saturated random coverage ratio for disks on a flat collector \cite{bib:Zhang2013,bib:Ciesla2013pol}.}
\label{fig:qmax}
\end{figure}

Having determined the value of exponent $d$, Eq.(\ref{fl}) could be transformed to linear relation $\theta(y) = \theta_{\rm max} - A y$, where $y=t^{-1/d}$ and $A$ is a constant coefficient. Approximation of the function for $y=0$ gives the saturated random coverage ratio presented in Fig.\ref{fig:qmax}.
For $\epsilon<0.01$, $\theta_{\rm max}$ for a generalized dimer is the same as for disks. For $\epsilon=0.5$, a slight increase up to $\theta_{\rm max}=0.5793$ can be observed. Then, the saturated coverage decreases because the amount of blocked surface grows faster than dimer area. The minimum is reached around $\epsilon=1.5$ when particles disks are separated, but the distance between them is too small to be filled up by another molecule. For $\epsilon=2$, it becomes possible, so the next local maximum is reached. For large $\epsilon$, saturated random coverage seems to approach again the value known for RSA of disks. 
\subsection{Available surface function}
\begin{figure}[htb]
\vspace{1cm}
\centerline{%
\includegraphics[width=0.8\columnwidth]{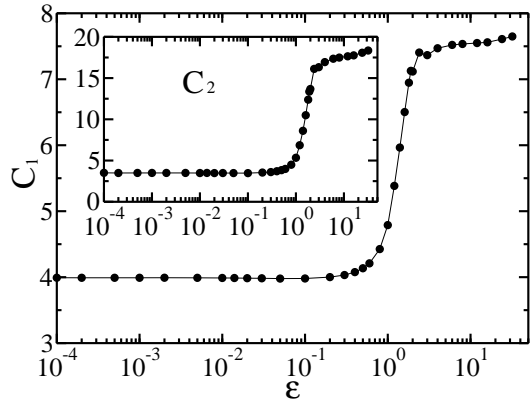}}
\caption{The dependence of $C_1$ and $C_2$ (inset) coefficients on dimer anisotropy $\epsilon$. Dots are simulation data. Solid lines have been drawn to guide the eye.}
\label{fig:c1c2}
\end{figure}

Available surface function (ASF) describes probability of placing subsequent particle on a collector as a function of  coverage, and therefore can be easily estimated using RSA simulations. It can be also measured experimentally \cite{bib:Schaaf1995,bib:Adamczyk1996}. ASF is important for at least two reasons. Firstly, together with the model of a specific transport mechanism, responsible for bringing particles to the close proximity of a collector surface, it allows to predict kinetics of the monolayer growth \cite{bib:Adamczyk2010,bib:Ciesla2013dim,bib:Adamczyk1987,bib:Erban2007a,bib:Erban2007b}. Secondly, the polynomial expansion of ASF for low $\theta$:
\begin{equation}
ASF(\theta) = 1 - C_1\theta + C_2\theta^2 + o(\theta^2)
\end{equation}  
is closely related to the viral expansion of the state equation of a formed monolayer. For example, second and third viral coefficients are $B_2=1/2C_1$ and $B_3=1/3C_1^2-2/3C_2$, respectively \cite{bib:Tarjus1991,  bib:AdamczykBook}. It is also worth noticing that coefficient $C_1$ corresponds to the surface area blocked by a single particle, whereas $C_2$ denotes a cross-section of the area blocked by two independent molecules. Their dependence on anisotropy parameter is shown in Fig.\ref{fig:c1c2}.
Just like for saturated coverage ratio, for small anisotropy, both the coefficient are equal to their values known from disks monolayers: $C_1^{disk} = 4$ and $C_2^{disk} = \frac{6\sqrt{3}}{\pi} \approx 3.308$. On the other hand, for a large distance between disks in the generalized dimer, it is expected that $C_1 \to 2 C_1^{disk} = 8$ and $C_2 = 2 \left( C_2^{disk} \right)^2 \approx 21.885$. Observed values in Fig.\ref{fig:c1c2} are slightly lower, probably due to not large enough $\epsilon$. Transition between those two limits occurs mainly between $\epsilon=1$ and $\epsilon=2$, which is directly connected with increasing probability of two different particles crossing.
\subsection{Density autocorrelation function}
\begin{figure}[htb]
\vspace{1cm}
\centerline{%
\includegraphics[width=0.6\columnwidth]{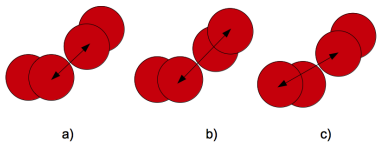}}
\caption{(Color online) Two exemplary generalized dimers ($\epsilon=0.5$) taken from saturated random coverage. Arrows points distances between disks from different particles. In case a) the distance is sligtly higher than $2r_0$ wheras in cases b) and c) it is equal about $3r_0$.}
\label{fig:g_exp}
\end{figure}

Saturated random coverage ratio describes mean density of adsorbed particles inside formed adsorption monolayer. More information about the monolayer structure can be acquired by analyzing density autocorrelation function, which is usually defined as a normalized probability of finding other particle in specified distance range:
\begin{equation}
G(r) = \frac{P(r)}{2\pi r \rho}.
\end{equation}
Here $P(r)dr$ is the probability of finding two disks in a distance between $r$ and $r+dr$, measured between their geometric centers.  Parameter $\rho$ is the mean density of particles inside a covering layer. Such a normalization leads to $G(r\to \infty) = 1$. In the case of spherical particles, $G(r)$ has a logarithmic singularity in the touching limit \cite{bib:Swendsen1981} and superexponential decay at large distances \cite{bib:Bonnier1994}. 
\begin{figure}[htb]
\vspace{1cm}
\centerline{%
\includegraphics[width=0.8\columnwidth]{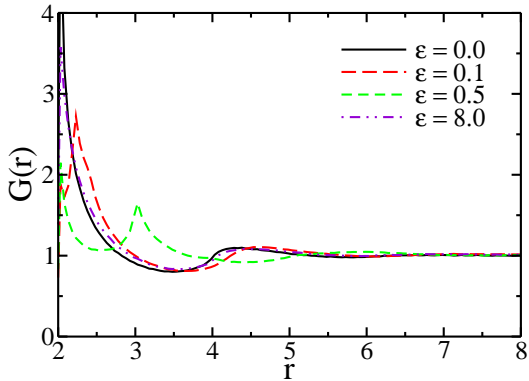}}
\caption{(Color online) Density autocorrelation function for several dimers of different anisotropy $\epsilon$. Distance (in $r_0$ units) is measured between centers of different disks, no matter whether they come from the same or different dimers. Autocorrelation function was measured for layers obtained at the end of simulations. Their coverage ratio was about $0.5\%$ smaller than $\theta_{\rm max}$.}
\label{fig:g}
\end{figure}
As shown in Fig.\ref{fig:g}, when disks in a single particle are well separated ($\epsilon=8$) the autocorrelation function is nearly the same as for ($\epsilon=0$). Moreover, all plots are almost the same for large enough $r$, however, local maxima, resulting from a fixed distance between disks in a single particle, can appear. To understand this phenomenon let's focus on one specific case, e.g. $\epsilon=0.5$. As the coverage is saturated closest particles are nearly touching themselves (see Fig.\ref{fig:g_exp}). 
Therefore, it is clear that distance between centers of touching disks is only slightly higher than $2r_0$ (Fig.\ref{fig:g_exp}a), and, as it is a common situation in saturated coverage, the density autocorrelation function will show a maximum there. However, the distance between the touching disk and the other one in second particle is most likely around $3r_0$ (Fig.\ref{fig:g_exp}b and \ref{fig:g_exp}c). Therefore, for $\epsilon=0.5$ the density autocorrelation function has another maximum at $r \approx 3r_0$. It is worth to notice that for small $\epsilon$ the first maximum at $r=2 r_0$ could be smaller than the second maximum (see Fig.\ref{fig:g} for $\epsilon=0.1$). On the other hand, for $\epsilon \ge 1$ there is also a maximum at $2\epsilon r_0$, which  corresponds to distance between disks in the same dimers.
\subsection{Random packing of generalized dimers in 3D}
\begin{figure}[htb]
\vspace{1cm}
\centerline{%
\includegraphics[width=0.8\columnwidth]{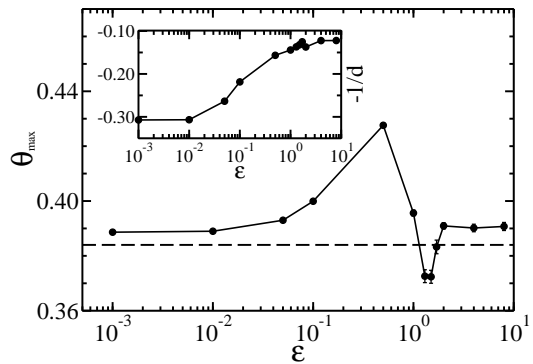}}
\caption{Saturated random coverage ratio measured for several dimers of different anisotropy $\epsilon$. The dashed line corresponds to $\theta_{\rm max} = 0.3841$, which is the saturated random coverage ratio of spheres in three dimensional space \cite{bib:Zhang2013}. The solid line is to guide the eye. Inset shows the exponent in Eq.(\ref{fl}) determined using least square fit method dependence on particle anisotropy.}
\label{fig:3d}
\end{figure}

As mentioned above, parameter $d$ in Eq.(\ref{fl}) corresponds to a number of degrees of freedom of adsorbate particle. Therefore in two dimensional case the transition between $d=2$ and $d=3$ has been observed. In three dimension a sphere has $d=3$ degrees of freedom, whereas dimer ($\epsilon=1$) is described by two additional angles, which determine its orientation ($d=5$). It is interesting to see if here RSA kinetics behaves similarly as in 2D case, and especially where the transition point is. To check this, additional simulations were performed. Generalized dimers build of two spheres of radius $r_0$ were placed randomly in a cubic box of side size $80 r_0$. The number of RSA algorithm steps in a single simulation expressed in dimensionless time units (\ref{eq:dimlesstime}) was equal to $10^5$, as in 2D case, but here
\begin{equation}
S_{\rm M}(\epsilon) = \left\{ 
  \begin{array}{cc} \frac{2}{3}\pi \left[ 4 - \left( 1-\epsilon \right)^2 \left(2+\epsilon \right) \right] r_0^3  & \mbox{if } \epsilon<1, \\
\frac{8}{3} \pi r_0^3 & \mbox{ otherwise},
\end{array}
\right.
\end{equation}
Results are presented in Fig.\ref{fig:3d}.
As expected, for small anisotropy RSA kinetics is governed by $d=3$ exponent. This changes around $\epsilon=0.05$ when the exponent starts to increase. It is worth to remind, than in two dimensional case this growth started at $\epsilon=0.01$. For large anisotropy ($\epsilon>2$), when spheres in a dimer are separated, parameter $d$ is significantly higher than molecules number of degree of freedom. The dependence of the saturated random coverage ratio on anisotropy $\epsilon$ is qualitatively same as in two-dimensional case except that there is no sharp maximum at $\epsilon=1$. The slight shift between measured coverage ratio and its values determined for spheres in \cite{bib:Zhang2013} are probably due to different boundary conditions. Here, the three dimensional cube containing molecules was fixed, whereas in \cite{bib:Zhang2013} periodic boundary conditions were used.
\section{Summary} 
The RSA kinetics of generalized dimers depends strongly on particle anisotropy. In two dimensional case, the exponent in Eq.\ref{fl} varies from $d=2$, known from spheres adsorption, to $d=4.5$, with surprisingly small level of anisotropy ($\epsilon=0.01$) inducing its increase to $d \approx 3.0$, typical for an anisotropic particle. In three dimensions exponent $d$ varies from $d=3$ for small anisotropy ($\epsilon \le 0.01$) to $d=6$ for $\epsilon>2$. The saturated random coverage ratio is similar to the one for disks for $\epsilon<0.02$ and $\epsilon >8$. For intermediate anisotropy, it reaches maximum $\theta_{\rm max} = 0.5793 \pm 0.0001$ for $\epsilon=0.5$ and minimum $\theta_{\rm max} = 0.4786$ around $\epsilon = 1.5$. In three dimensions, results are qualitatively same with $\theta_{\rm max}$ varying from $0.373$ for $\epsilon \approx 1.6$ to $0.428$ for $\epsilon \approx 0.5$. Viral expansion coefficient for small and large $\epsilon$ are the same as for disks, keeping in mind that for large $\epsilon$, single particle contains two of them. The transition between these two limits takes place for $\epsilon \in (1, 2)$. The density autocorrelation behaves similarly as in the case of disks; however, for small $\epsilon$, its plot is shifted accordingly to the displacement of disks within the dimer.
\section*{Acknowledgments}
I would like to thank Jakub Barbasz for many fruitful discussions and helpful comments.
This work was supported by Polish National Science Center grant no. UMO-2012/07/B/ST4/00559.
%

\end{document}